# Interlayer-Exciton Based Nonvolatile Valleytronic Memory

*Tong Ye[1], Yongzhuo Li[3,4], Junze Li[1], Hongzhi Shen[1], Junwen Ren[1], Cun-Zheng Ning[3,4,5], Dehui Li[1,2]\**

[1]School of Optical and Electronic Information, Huazhong University of Science and Technology, Wuhan 430074, China.

[2]Wuhan National Laboratory for Optoelectronics, Huazhong University of Science and Technology, Wuhan 430074, China.

[3]Department of Electronic Engineering, Tsinghua University, 100084 Beijing, China.

[4]Frontier Science Center for Quantum Information, 100084 Beijing, China.

[5]School of Electrical, Computer, and Energy Engineering, Arizona State University, Tempe, AZ 85287, USA.

**Analogous to conventional charge-based electronics, valleytronics aims at encoding data via the valley degree of freedom, enabling new routes for information processing. Long-lived interlayer excitons (IXs) in van der Waals heterostructures (HSs) stacked by transition metal dichalcogenides (TMDs) carry valley-polarized information and thus could find promising applications in valleytronic devices. Although great progress of studies on valleytronic devices has been achieved, nonvolatile valleytronic memory, an indispensable device in**

**valleytronics, is still lacking up to date. Here, we demonstrate an IX-based nonvolatile valleytronic memory in a $WS_2/WSe_2$ HS. In this device, the emission characteristics of IXs exhibit a large excitonic/valleytronic hysteresis upon cyclic-voltage sweeping, which is ascribed to the chemical-doping of $O_2/H_2O$ redox couple trapped between the TMDs and substrate. Taking advantage of the large hysteresis, the first nonvolatile valleytronic memory has been successfully made, which shows a good performance with retention time exceeding 60 minutes. These findings open up an avenue for nonvolatile valleytronic memory and could stimulate more investigations on valleytronic devices.**

Van der Waals HSs stacked by TMDs monolayers enable the generation of long-lived IXs with a large binding energy of about 150 meV[1], and a long diffusion distance over five micrometres[2], further extending the already appealing properties of the constituent TMDs monolayers. Since IXs are composed of electrons and holes that are resided in neighboring layers, their physical properties strongly depend on the layer configurations and external fields or dopings[3,4]. Through electrical field or doping, we can modulate the emission intensity and wavelength of the IXs[1], and even switch its polarization[5]. Recently, IXs in the HSs stacked by other layered materials such as 2D perovskites and InSe with TMDs monolayer have been demonstrated and can be utilized in mid-infrared photodetections[6,7].

In particular, IXs in TMDs-based heterostructures carry valley-polarized information and thus would find promising applications in valleytronics taking advantage of their long lifetime[8]. Previous studies have demonstrated that IXs exhibit a large valley-polarization degree that can be tuned in a wide range by external electric field[9] and magnetic field[10]. Although considerable progress has been made in valleytronics, nonvolatile valleytronic memory has not been achieved up to date, which is indispensable for valleytronic devices. To this end, it is urgent to explore possible strategies to efficiently store valley-polarized information for further development of valleytronic devices and chips. Here, we have successfully achieved an IX-based nonvolatile valleytronic memory, which would greatly prompt relevant investigations on valleytronics.

In this work, the HS device is formed by a monolayer $WS_2$ (top) and a monolayer $WSe_2$ (bottom), both of which are contacted with an electrode (Fig. 1a). By applying voltage between the electrode and the heavily-doped Si substrate, we can control the doping level of the device when performing optical measurements. Figure 1b shows the optical microscope image of the device. The $WS_2$ and $WSe_2$ sheets are mechanically exfoliated from their respective bulk crystals and then transferred on a $SiO_2/p^{++}$-Si substrate through dry-transfer technique[11]. The edges of the two sheets are intentionally aligned to improve interlayer coupling[3].

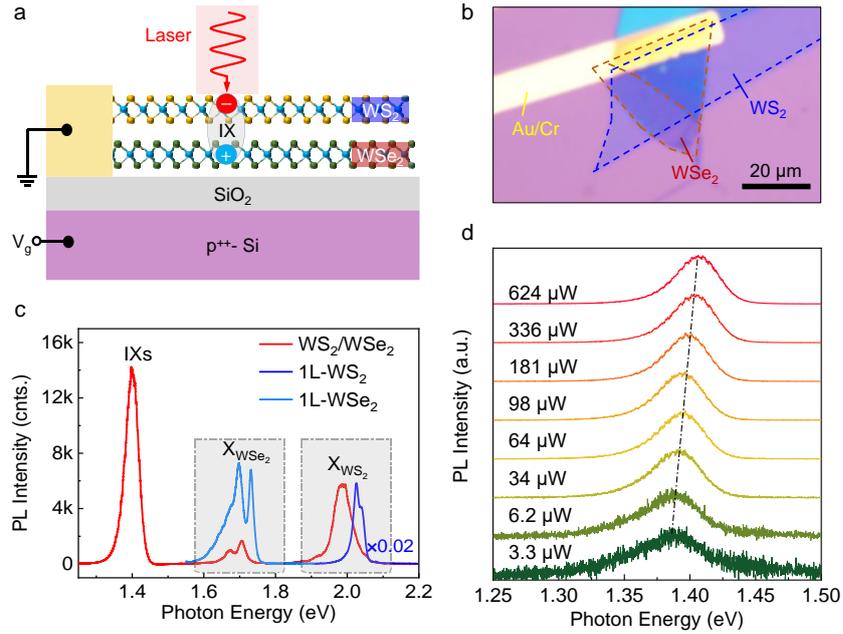

**Fig. 1|IXs in a WS$_2$/WSe$_2$ HS. a-b**, Schematic and optical microscope image of the device, respectively. **c**, PL spectra of the HS and monolayer WSe$_2$ and WS$_2$. For a clear visualization, the PL spectrum of individual WS$_2$ is multiplied by a factor 0.02. **d**, PL spectra of the IXs as a function of excitation power. The spectra are vertically shifted for clarity. The sample was excited by a 532 nm laser with a power of 23 μW at 78 K.

**IXs in the WS$_2$/WSe$_2$ HS.** Fig. 1c shows the PL spectra of the HS, from which we can observe severe PL quenching and redshift of the intralayer excitonic peaks, together with the appearance of a low-energy peak at 1.4 eV. The quenching and redshift of the intralayer excitonic peaks can be attributed to interlayer charge transfer[12,13] and modified dielectric environment[14,15], respectively. We ascribe the peak at 1.4 eV to the IX emission according to previous reports[16,17]. The excitation-power dependent PL spectra further verify its interlayer nature (Fig. 1d). The IX emission peak shows a blueshift with the increase of excitation power, which is due to many-body effect arising from the repulsive interaction between the dipole-aligned IXs[4,18,19].

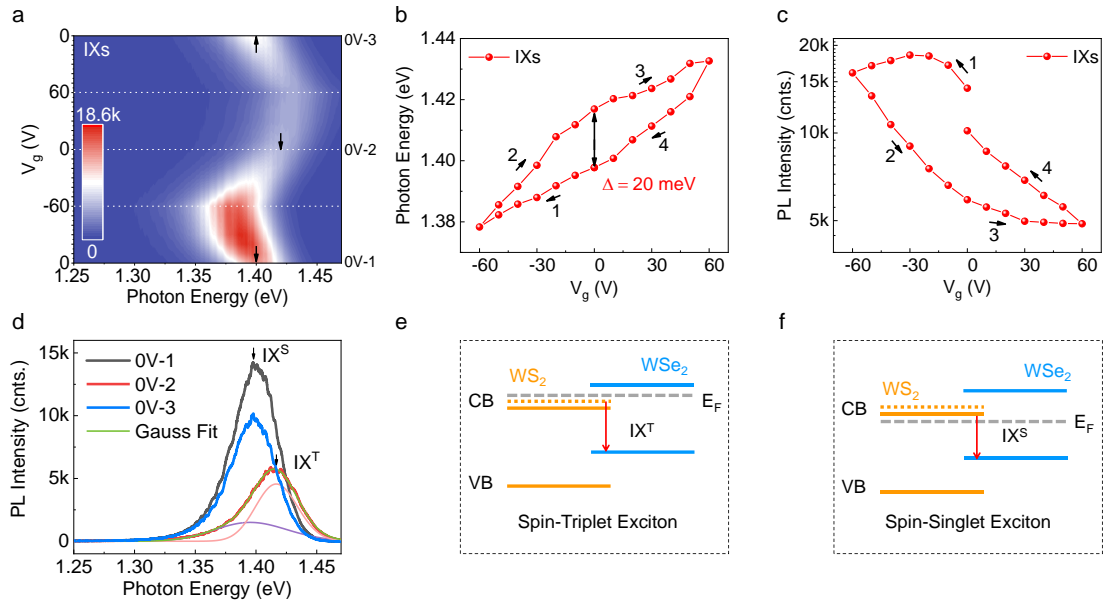

**Fig. 2|Electrical control of the IX emission. a**, Contour plot for the PL spectra of IXs upon cyclic $V_g$. The white dashed lines serve as guides to the eye. Black arrows mark the peak positions of the IXs at 0 V with different scanning sequences. **b-c**, Photon energy and PL intensity of the IX emission as a function of $V_g$. **d**, PL spectra of the IXs at 0 V with different scanning sequences. 0V-1, 0V-2 and 0V-3 represent three spectra marked in **a**. The 0V-2 spectrum is fitted by a Gaussian function. The sample was excited by a 532 nm laser with 23 μW power at 78 K. **e-f**, Schematic of the spin-triplet and spin-singlet excitons. Chemical-doped electrons lift the Fermi level up and shift the IXs to the spin-triplet state ($IX^T$). When those electrons are released, the IXs return to the spin-singlet state ($IX^S$). The orange dashed line stands for the upper spin-splitting conduction band (CB) of $WS_2$. Red arrows represent the recombination paths of the IXs.

**Excitonic hysteresis of IXs.** To explore gate-dependent features of the IX emission, we measured the PL spectra of the device under cyclic $V_g$, which scans first from 0 V to −60 V, then 0 V all the way to 60 V and finally back to 0 V (Fig. 2a). The IX emission peak shows a redshift and the emission intensity is enhanced with the decrease of $V_g$, and vice versa. The redshift of the IX emission peak with $V_g$ can be ascribed to the Stark effect[5,20], which is further verified by the opposite shift trend of the IX emission peak in devices with stacking order inversed (Fig. S1). Interestingly, the IX emission peak exhibits a strong hysteresis upon cyclic-voltage sweeping. As indicated by the black arrows in Fig. 2a, the peak energy of the IXs at middle 0 V (0V-2) cannot return to the same value of initial 0 V (0V-1), until a further upward scanning that is finally back to 0 V (0V-3). The gate-dependent photon energy and PL intensity can be seen more clearly in Fig. 2b and 2c. For a simple discussion, we only compare the states at 0V-2 and 0V-3. The photon energy of 0V-2 is blueshifted by about 20 meV with respect to that of 0V-3. Meanwhile, the PL intensity of 0V-2 is weaker than 0V-3 with a contrast ratio of about 1.7. It is worth to mention that the light intensity changes non-monotonously as $V_g$ decreases from 0 V to −60 V, indicating the occurrence of chemical doping[21-23], which will be discussed in the following.

As shown in Fig. 2d, the IX emission peak of 0V-2 can be decomposed to two Gaussian peaks (detailed fittings of the spectra are provided in Fig. S2). The energy difference of the two peaks is about 20 meV, which is consistent with the splitting energy of the conduction band of $WS_2$[24,25], strongly suggesting the occurrence of spin-triplet excitons[26,27]. This peculiar phenomenon can be understood from the chemical-doping[21,22] induced band-filling effect[5,28], as depicted in Fig. 2e and 2f. When the device is chemically $n$-doped, the Fermi level will be lift up and IXs will shift to the spin-triplet state (IX$^T$), which has an inefficient PL yield because of inversed spin. Contrarily, when the chemically-doped electrons are released, IXs will return to the spin-singlet state (IX$^S$). Therefore, the IX emission peaks in 0V-1 and 0V-3 spectra are attributed to IX$^S$ emission, and that in 0V-2 spectra is mainly resulted from IX$^T$. The IX$^T$ and IX$^S$ peaks can be well resolved in PL spectra acquired by picosecond laser excitation (Fig. S3a). In addition, the intensity ratio of IX$^T$/IX$^S$ increases with the increase of $V_g$ (Fig. S3b), thus confirming the band-filling mechanism and IX$^T$/IX$^S$ origins. We have also measured the gate-dependent lifetime of the IXs (Fig. S3, c-e). The lifetime of the IXs at 0V-2 is slightly shorter than at 0V-1 and 0V-3 rather than getting prolonged, further supporting the IX$^T$/IX$^S$ origins[10]. The difference of light intensities between 0V-3 and 0V-1 might be due to different levels of chemical doping at the initial and final sweeping stages, as discussed in the following.

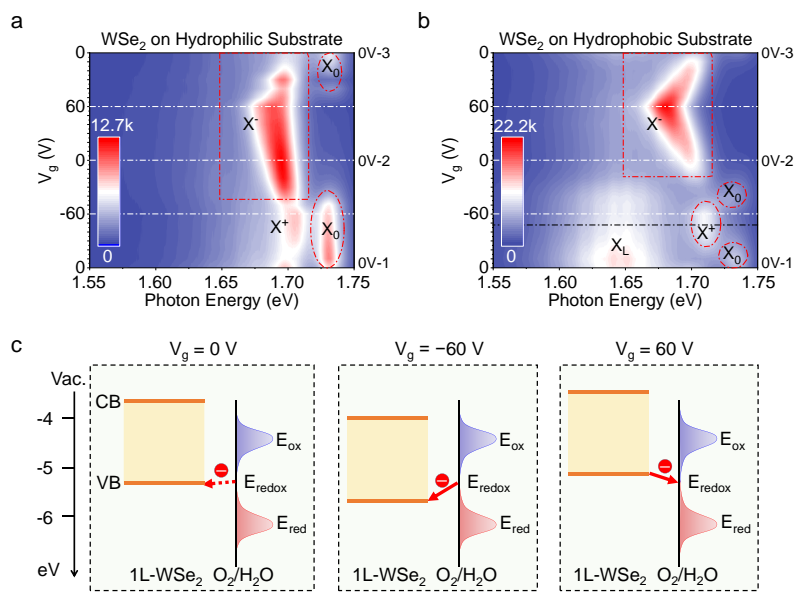

**Fig. 3|Mechanism of the excitonic hysteresis. (a)** Contour plot for the PL spectra of monolayer WSe$_2$ as a function of cyclic $V_g$. The spectra were acquired in the individual WSe$_2$ region of the HS on a hydrophilic substrate. **(b)** Contour plot for the PL spectra of

monolayer WSe$_2$ on a hydrophobic substrate, which is functionalized by hexamethyldisilazane (HMDS). The black dashed line indicates the symmetric position of the evolution tracks of $X^+$ and $X_0$. The PL measurements were conducted at 78 K with 532 nm laser excitation (23 μW). **(c)** Illustration of chemical doping caused by O$_2$/H$_2$O molecules. The electronic density of states (DOS) reflect the electron energy distribution around the oxidation potential (E$_{ox}$) and reduction potential (E$_{red}$), respectively.

**Mechanism of the excitonic hysteresis.** Electrical hysteresis has been observed in devices based on two-dimensional materials, such as graphene and TMDs based field-effect transistors[29,30]. Generally, electrical hysteresis is attributed to the chemical-doping effect by doping species (O$_2$ and H$_2$O) that are bound at the device/substrate interface, and/or on the surface of the device[31-33]. In our case, we propose that the excitonic hysteresis mentioned above is originated from the same scenario.

Since our measurements were performed in high vacuum (~10$^{-7}$ Torr), the influence of the molecules on the device surface can be safely neglected. Therefore, the excitonic hysteresis is more likely due to the O$_2$/H$_2$O molecules that are trapped at the interface between the HS and substrate. To clarify this, we examine the gate-dependent PL spectra of the individual WSe$_2$ region (Fig. 3a), because WSe$_2$ is in the bottom of the HS and directly contacts the SiO$_2$/Si substrate. Additionally, we conducted a control experiment with WSe$_2$ monolayer on a hydrophobic substrate (Fig. 3b).

The emission features of the intralayer excitons in WSe$_2$ are closely correlated to that of IXs. As $V_g$ decreases from 0 V to –60 V, the emission of positive trions ($X^+$) is gradually enhanced, while the peak of neutral excitons ($X_0$) is suppressed, indicating an efficient hole doping (detailed data is provided in Fig. S4). Peculiarly, as $V_g$ increases from −60 V back to 0 V, the evolution track is asymmetric to that from 0 V to –60 V. The trion emission peak is firstly weakened, then enhanced and redshifted with the increase of $V_g$. The asymmetric evolution strongly indicates the occurrence of negative trions ($X^-$), and suggests that the WSe$_2$ is chemically $n$-doped[33,34] at 0V-2. When $V_g$ increases from 0 V to 60 V, the $X^-$ peak is redshifted further, but with emission intensity weakened because of Coulomb screening from the free electrons[35]. When voltage scans backward from 60 V to 0 V, the $X^-$ peak shows a blueshift and the emission intensity becomes weaker while the $X_0$ peak is gradually enhanced, indicating that the chemically-doped electrons have been released. All the above features are well consistent with the previously mentioned chemical-doping effect.

    To further validate such hypothesis, we then focus on the PL spectra of a control device with monolayer WSe$_2$ on a hydrophobic substrate (Fig. 3b). The evolution tracks of $X^+$ and $X_0$ emission are roughly symmetrical along the black dashed line at about –50 V. The slight deviation of the symmetry line at –50 V (rather than –60 V) might be due to trace O$_2$/H$_2$O molecules that are adsorbed on WSe$_2$ before the transfer procedure. Besides, in sharp contrast to Fig. 3a, the track of $X^-$ is quasi-symmetrical along the dashed line at 60 V, suggesting that the excitonic hysteresis is largely suppressed. Therefore, H$_2$O molecules should play a critical role in our observations. The broad PL peak centered at about 1.65 eV might be due to local-state exciton ($X_L$)[36], which is out of

the scope of this study.

The surface of $SiO_2$ is usually covered with a layer of silanol groups ($\equiv Si - OH$), especially after it is treated by piranha solution or plasma cleaner[21,33]. With these silanol groups, $SiO_2$/Si substrates are easily bound by ambient $O_2$ and $H_2O$ molecules[33]. As shown in Fig. 3c, the electrochemical potential of the redox couple ($O_2/H_2O$) is about −5.3 eV[21,37], which is slightly higher than the valence band of $WSe_2$ (about −5.46 eV)[38,39]. Therefore, electrons spontaneously transfer from $O_2/H_2O$ to $WSe_2$, making monolayer $WSe_2$ initially *n*-doped (detailed information is provided in Fig. S4), and resulting in the deviated symmetry at −50 V in Fig. 3b.

When applying negative gate voltages, electrons are forced to transfer further from $O_2/H_2O$ to $WSe_2$. Consequently, the Fermi level of the HS is lifted up, and IXs shift to the spin-triplet state (Fig. 2e). Those chemically-doped electrons balance out the gate modulation, resulting in the non-monotonic behavior of the IXs in 0 ∼ −60 V range (Fig. 2c) and the excitonic hysteresis. The chemical-doping effect also explains why $X_0$ emission maintains its intensity from 0 V to −60 V for $WSe_2$ on the hydrophilic substrate (Fig. 3a) but greatly suppressed on the hydrophobic substrate (Fig. 3b).

When applying positive gate voltages, the chemical-doped electrons are driven back from $WSe_2$ to the $O_2/H_2O$ redox couple. Therefore, IXs return to the spin-singlet state (Fig. 2f) when $V_g$ scans back to 0V-3. This control experiment further verifies the chemical-doping mechanism and well explains the origin of the excitonic hysteresis of IXs shown in Fig. 2. We have made dozens of HSs on hydrophobic and hydrophilic substrates, and we can only observe excitonic hysteresis (both photon-energy and PL-intensity hysteresis) in the samples on hydrophilic substrates. The hysteresis is largely suppressed in HSs stacked on hydrophobic substrates (Fig. S5). In addition, we have also fabricated a $WS_2/WSe_2/hBN$ HS on a hydrophilic substrate with $WS_2/WSe_2$ HS partially separated from the substrate by a thin layer hBN. For this device, the excitonic hysteresis is observed in the region where $WS_2/WSe_2$ HS directly contacts with the substrate, but absent in the hBN-insulated region (Fig. S6), further supporting the chemical-doping mechanism. The hysteretic behavior is well reproducible in multiple repeating measurements and also in different samples. Therefore, we rule out the influence of random contamination.

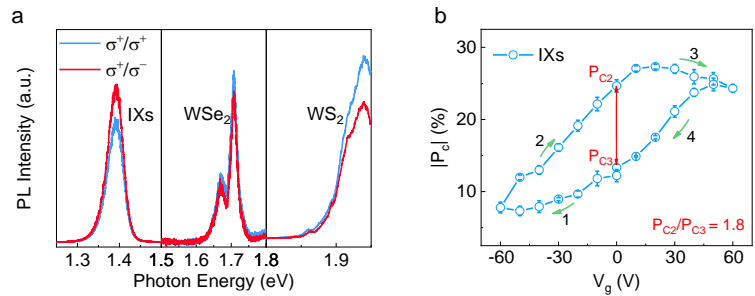

**Fig. 4|Electrically-tunable valley polarization of the IXs. a,** Helicity-resolved PL spectra of the HS under 633 nm excitation (180 μW) at 78 K. **b,** Absolute circular polarization degree of the IXs as a function of $V_g$. The helicity contrast is defined as $P_{c2}/P_{c3}$, where $P_{c2}$ and $P_{c3}$ is the absolute circular polarization degree of 0V-2 and 0V-3, respectively.

**Valleytronic hysteresis of the IXs.** To study the chemical-doping effect on the valley-polarized features of the IXs, we measured the helicity-resolved PL spectra of the device (Fig. 4a). Interestingly, the IX peak exhibits a negative circular polarization in contrast to that of intralayer excitons in $WSe_2$ and $WS_2$, which can be ascribed to the interlayer quantum interference imposed by the atomic registry between the constituent layers[40]. To qualify the valley polarization, the degree of circular polarization (DOCP) has been introduced and defined as $P_c = (I^+ - I^-)/(I^+ + I^-)$, where $I^+$ ($I^-$) denotes the intensity of co-polarized (cross-polarized) PL component. For the IXs peak, $P_c = -12.3\%$, while for the intralayer excitonic peak of WS$_2$ and WSe$_2$, $P_c = 15\%$ and 7.1%, respectively.

The DOCP of the IXs can also be electrically controlled by $V_g$, as shown in Fig. 4b (the full data set is provided in Fig. S7). The absolute DOCP is greatly suppressed at −60 V (*p*-doping), but enhanced at 60 V (*n*-doping). This phenomenon has been reported by Scuri and coworkers, and is attributed to changes in valley-depolarization-time caused by electron/hole doping[41]. Similarly, we believe our observation can also ascribed to the charge doping from external applied bias and chemical doping (Fig. S3). Interestingly, the DOCP and lifetime (Fig. S3e) of the IXs also exhibit a strong hysteresis, probably due to the carrier trapping and detrapping induced by the above-mentioned chemical-doping, which leads to the different doping concentration and further different valley-depolarization-time and DOCP under the same gate voltage. To sum up, the chemical-doping effect leads to the formation of spin-triplet exciton, and gives rise to the hysteresis of excitonic emission, valley-polarization degree and lifetime of IXs, which could find potential applications in nonvolatile valleytronic information processing.

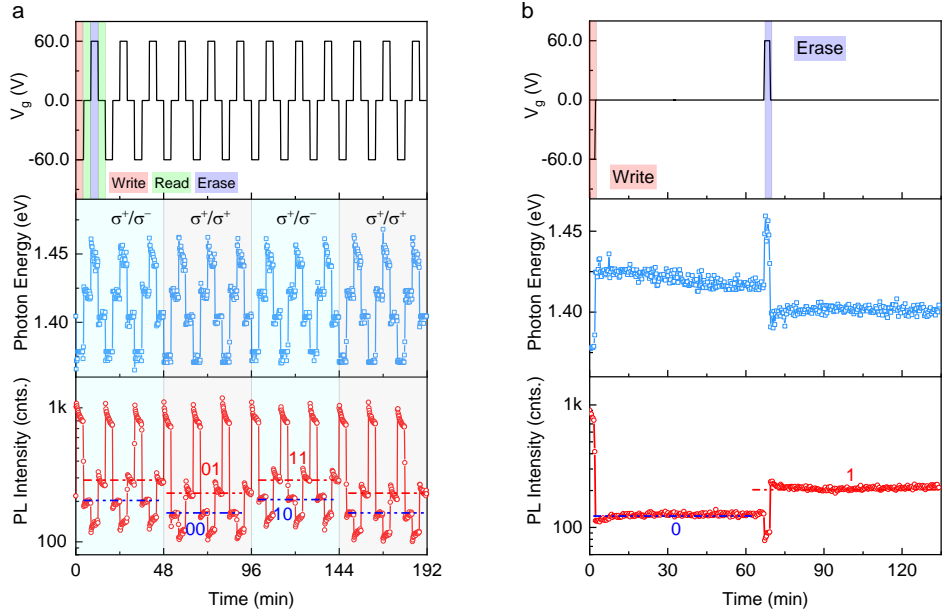

**Fig. 5 | Electrically controlled memory operation in the HS. a,** Time-dependent IX emission characteristics upon cyclic $V_g$ among −60 V, 0 V, 60 V and 0 V. Each voltage lasts for about 4 minutes. The detecting polarization shifts every three cycles of $V_g$. **b,** Retention time of the "1" and "0" excitonic states. The writing and erasing voltages last for about 3 minutes, and the reading voltage lasts for about 64 minutes. The peak energies and intensities are extracted from real-time spectra, each of which was measured within 10 s. The sample was excited by a 633 nm laser with a power of 180 μW at 78 K.

**IX-based valleytronic memory.** To demonstrate the valleytronic memory ability of the device, we measured time-dependent PL spectra under circular excitation ($\sigma^+$), as shown in Fig. 5a. As gate voltage cyclically changes among −60 V, 0 V, 60 V and 0 V, the photon energy of the IX emission periodically shifts among 1.38 eV, 1.42 eV, 1.45 eV and 1.40 eV, which are analogous to the performance of conventional electronic devices under "write", "read" and "erase" operations. In addition, the emission intensity also periodically changes in response to those memory operations. Specifically, the intensity level of $IX^S$ ($IX^T$) located at 1.40 (1.42) eV can be regarded as digital information 1 (0), which can persist for a long time with no power consumption, suggesting potential applications in nonvolatile storage. Intriguingly, as the detection helicity switches between $\sigma^-$ and $\sigma^+$, the PL intensity of the 0 and 1 states exhibit helicity-resolved features. There are four intensity levels emerging, which can be defined as "00", "01", "10"and "11", indicating valley-encoding abilities of the device. Based on this feature, we can selectively encode/address the valley-polarized information by helicity excitation/detection.

To evaluate the retention time of the encoded information, we then prolong the reading-operation time, as shown in Fig. 5b. Surprisingly, the 1 and 0 excitonic states can persist for at least 60 minutes, holding great promise for nonvolatile valleytronic memories. As a matter of fact, the retention time should be much longer than 60 minutes, as can be seen in a logarithmic-timescale plot (Fig. S8a). We also note that the 0 (1) state varies dynamically before reaching a steady state. This is probably due to the charging/discharging process of the device, as confirmed by the features of time-dependent leakage current (Fig. S8b). It is worth to mention that the information encoding ability of the device can persist up to 250 K, which is promising for high temperature valleytronic applications (Fig. S9).

Since the nonvolatile valleytronic memory has never been reported, it is hard to make an objective comparison. Nevertheless, the device is similar to photonic memory, thus we list the parameters of our device and other nonvolatile photonic memories in

Table 1, which shows that our device is outperforming in comparison with peer memory devices. The PL ON/OFF ratio of the 1/0 states could be as large as 3.6 (Fig. S10), which is larger than peer photonic memories[42-45]. The power consumption of the device is estimated to be about 74/56 nW for set/reset operation (Fig. S8b), which is extremely low in comparison with other phase-change photonic memories[42-45]. The switching time of our devices could be very short but limited by our testing system, since the hysteresis effect could be established in several microseconds according to previous reports[46].

**Table 1.** Parameters of our device and peer works. The abbreviations of ele-photonic, E.P. and O.P. stand for electrical-photonic, electrical programing, and optical programing, respectively.

| Memory Type | ON/OFF Ratio | Operation Time (ns) | Power(Set/Reset) (mW) | Ref. |
|---|---|---|---|---|
| all-photonic | 1.21 | 1 | 53.3 (O.P.) | [42] |
| all-photonic | 1.8 | 5 | 10/30 (O.P.) | [45] |
| ele-photonic | 3.16 | 80100 | 10/110 (E.P.) | [43] |
| ele-photonic | 1.04 | 510 (E.P.) 408 (O.P.) | 0.03/1.2 (E.P.) 7.5 (O.P.) | [44] |
| ele-photonic | **3.6** | -- | 74/56 *$10^{-6}$ (E.P.) | This work |

CONCLUSIONS

In summary, we have systematically investigated the excitonic/valleytronic hysteresis of IXs in a $WS_2/WSe_2$ HS. By examining the PL spectra of the $WSe_2$ monolayers on hydrophilic and hydrophobic substrates, we verify that the origin of the hysteresis is the chemical-doping of WSe$_2$ by O$_2$/H$_2$O redox couple. Benefiting from the hysteresis effect, IXs can be electrically switched between a spin-singlet state and a spin-triplet state, enabling the applications in valleytronic information processing. Finally, we demonstrate the memory function of the device, which shows a good writing/reading/erasing ability with retention time exceeding 60 minutes. Our study provides a potential paradigm to achieve nonvolatile valleytronic memory and thus would greatly advance the development of valleytronic devices.

**METHODS**

***Sample Preparations.*** Electrodes were fabricated by standard photolithography and thermal evaporation (50 nm/2 nm Au/Cr). The substrates with prefabricated electrodes are ultrasonic cleaned and plasma cleaned before the fabrication of the HS. WS$_2$ and WSe$_2$ monolayer flakes were first mechanically exfoliated onto polymethyl-methacrylate (PMMA) stamps, and then transferred on a SiO$_2$ (300 nm)/Si wafer using a dry transfer technique with the aid of an optical microscope and a nano-manipulator. The hydrophobic substrates were prepared via immersing in HMDS vapor for 10 min and then rinsing with acetone for 30 s to form a hydrophobic layer on the substrate[47]. All the samples are not treated by thermal annealing, because the thermal-annealing procedure can disable or deteriorate the performance of nonvolatile memory devices.

***Optical Measurements***. The as-fabricated devices were mounted in a continuous flow cryostat with 10$^{-7}$ Torr vacuum. For gate-dependent PL measurement, the sample was excited by a 532 nm laser (23 μW) at 78 K. For the helicity-resolved PL measurement, the sample was excited by a 633 nm laser with a power of 180 μW at 78 K. The time interval between two adjacent spectra is about 1 minute when performing gate-dependent measurement. For the memory operation measurement, the spectra were acquired with $V_g$ changing cyclically and laser keeping focused on the sample. Each spectrum was measured within 10 seconds. All the PL spectra were collected by a 50× objective lens (N.A. = 0.7) in a Raman spectrometer (Horiba HR550) with a 600 g/mm grating. A Keithley 2400 sourcemeter was used as the voltage source.

## ASSOCIATED CONTENT

**Supporting Information**

The supporting information consists of: 1) Gate-dependent PL spectra for a stacking-order inversed HS ($WSe_2/WS_2$) on a hydrophilic substrate; 2)Lorentz and Gaussian fittings of 0V-1, V-2 and 0V-3 spectra; 3) Spin-singlet and spin-triplet IXs and their lifetimes; 4) PL spectra of the $WSe_2$ monolayer under gate voltages from 0 V to −60 V shown in Fig. 3a; 5) Gate-dependent PL spectra for a $WS_2/WSe_2$ HS on a hydrophobic substrate; 6) Gate-dependent PL spectra of a $WS_2/WSe_2/hBN$ HS; 7) Helicity–resolved PL spectra of the $WS_2/WSe_2$ HS; 8) Electrically controlled memory operation in the HS; 9) Memory performance of the $WS_2/WSe_2$ HS under different temperatures; 10) PL spectra of 0V-2 and 0V-3 states and the ON/OFF intensity Ratio.

## AUTHOR INFORMATION

**Corresponding Author**

*E-mail: dehuili@hust.edu.cn.

**Notes**

The authors declare no competing financial interest.

**ACKNOWLEDGMENT**

D. L. acknowledges the support from National Key Research and Development Program of China (2018YFA0704403), NSFC (62074064) and Innovation Fund of WNLO. T. Y. gratefully acknowledges, Jian Zhang, Danyang Zhang and Jiaqi Wang for the help in conducting experiments.